\documentclass[preprint,aps,12pt,preprintnumbers,eqsecnum,nofootinbib,superscriptaddress]{revtex4}
\usepackage{tabularx} % extra features for tabular environment
\usepackage{amsmath}  % improve math presentation
\usepackage{dsfont}
\usepackage{graphicx} % takes care of graphic including machinery
\usepackage[margin=1in,letterpaper]{geometry} % this shaves off default margins which are too big
\usepackage{physics}
\usepackage{ amssymb }
\usepackage[final]{hyperref} % adds hyper links inside the generated pdf file
\usepackage{simplewick}
\hypersetup{
	colorlinks=true,       % false: boxed links; true: colored links
	linkcolor=blue,        % color of internal links
	citecolor=blue,        % color of links to bibliography
	filecolor=magenta,     % color of file links
	urlcolor=blue         
}
\begin{document}

\title{\vspace*{0.5in} 
Predictions of Stochastic Composite Gravity\vskip 0.1in}
\author{Joshua Erlich}\email[]{erlich@physics.wm.edu}\affiliation{Department of Physics,
William \& Mary, P.O. Box 8795, Williamsburg, VA 23187-8795, USA}
\date{\today}

\begin{abstract}

In this essay I describe some new results of a framework for composite gravity coupled to matter. These include the Bekenstein-Hawking entropy formula, modifications to the low-$\ell$ moments of the CMB power spectrum, 
and new perspectives on the Hartle-Hawking no-boundary proposal and the initial conditions for inflation. We conclude with suggestions for experimental tests of the framework.

\vfill {\em Essay written for the Gravity Research Foundation 2025 Awards for Essays on Gravitation.}
\end{abstract}

\maketitle
\thispagestyle{empty}
%%%%%%%%%%%%%%%%%%%%%%%%%%%%%%%%%%%%%%%%%%%%%%%%%%%%%%%
\clearpage
\pagenumbering{arabic}
\section{Introduction}
The goal of this essay is to consider a foundation for the composite gravity paradigm and present some results and speculation. In stochastic composite gravity, both general relativity and quantum field theory are emergent descriptions of fluctuations in the equilibrium of a stochastic process that obeys certain dynamical equations. The basic framework was proposed in earlier work \cite{Erlich:2018qfc,Erlich:2024ooy}. The main novelty of this essay lies in the microscopic description of the discrete stochastic processes representing quantum states, discussion of the Hamiltonian and momentum constraints in the stochastic description, and presentation of some results and ideas, including:
\begin{itemize}
\item The emergence of a semiclassical spacetime consistent with Einstein's equations
\item The origin of gravitation from fluctuations about the semiclassical background
\item The Bekenstein-Hawking entropy formula
\item Consequences for cosmology
\item Ideas for experimental tests of the framework
\end{itemize}

\section{The Problem}
Lorentzian quantum field theory (QFT) and general relativity provide accurate descriptions of a wide range of observable and experimentally testable phenomena. However, combining the two frameworks into a coherent description of nature remains a challenge, with perturbative string theory and the AdS/CFT correspondence being the most well-studied attempts. Among the obstacles to formulating a quantum theory of gravity are the ontological constraints of quantum field theory and general relativity.

Quantum field theory is a probabilistic framework that includes in its axioms a Hilbert space of states with positive norm and an algebra of observables that act on the Hilbert space. Central to the framework is a notion of unitary time evolution. Unitarity ensures the conservation of probabilities as determined by the Born rule, and a Hamiltonian bounded below ensures stability of the time evolution.  

General relativity, on the other hand, is a theory for the geometry of spacetime and its interdependence on any other dynamical degrees of freedom. Dynamics is of a different character than in other physical theories: the spacetime geometry does not itself contain a definitive notion of time evolution, as a globally hyperbolic 4-manifold may be foliated by spacelike 3-manifolds in an infinititude of locally inequivalent ways. Each spacelike 3-dimensional leaf in the foliation has a distinct future and past,
but time evolution is a secondary construct, and one might argue an unnecessary one. 

The tension between the rigid notion of time in the quantum framework and the arbitrary parametrization of spacetime in general relativity leads to several conceptual puzzles, including the identification of physical frame-independent observables \cite{Giddings:2005id,Donnelly:2015hta}
and the interpretation of the timeless wavefunctional of the universe \cite{Page:1983uc,Isham:1992ms}. Puzzles related to unitarity and general covariance in black hole backgrounds add to the confusing state of affairs \cite{Hawking:1976ra,Almheiri:2012rt}.

Other questions arise from the holographic nature of gravity. 
The Bekenstein-Hawking formula for black hole entropy is given by \begin{equation}
S=\frac{c^3A}{4G\hbar}, \label{eq:Bek-Hawk}\end{equation}
where $A$ is the area of the black-hole horizon, and $G$ is Newton's gravitational constant. 
The microscopic interpretation of black-hole entropy remains a subject of debate. 

\section{A Framework for Quantum Gravity}
 
The challenge of reconciling the conflicting ontologies of quantum field theory and general relativity suggests that one or more basic premises of these frameworks may require revision. There are two complementary perspectives one may take: a bottom-up perspective in which we ask what new physics might appear in a regime of strong gravity, typically associated with the Planck scale $\ell_{\rm Pl}\sim10^{-35}$~m; and a top-down perspective in which QFT and/or general relativity are replaced by a new unifying framework from which standard physics is supposed to emerge. Both perspectives provide valuable insight.

Sakharov famously observed that the Einstein-Hilbert action of general relativity is induced in the effective action of a generic QFT in curved spacetime, as long as physics is modified at short distances so as to eliminate ultraviolet field-theory divergences \cite{Sakharov:1967pk}. Sakharov's observation suggests the ubiquity of an emergent gravitational interaction in background-independent frameworks that regularize QFT, should such frameworks exist at all. 
In this scenario, a semiclassical spacetime metric is identified with the expectation value of a composite operator determined by demanding that the energy-momentum tensor vanish.

Composite gravity models have received three main criticisms, which I would like to address before presenting the stochastic composite gravity framework:

1) The Weinberg-Witten theorem  is sometimes misunderstood to preclude the possibility of a massless composite spin-two state. The Weinberg-Witten theorem is obtained by considering rotation properties of matrix elements of a conserved Lorentz-covariant energy-momentum tensor between massless particle states \cite{Weinberg:1980kq}. 
However, in a coordinate-reparametrization-invariant setting, matrix elements of the energy-momentum tensor vanish and the Weinberg-Witten theorem does not apply.

2) It is sometimes argued that in Sakharov's induced gravity scenario, it is {\em ad hoc} for the auxiliary metric not to be dynamical from the outset \cite{Witten:2024upt}. From a bottom-up perspective this attitude is natural: Why should a gravitational interaction present at long distances vanish at some short-distance scale? However, in a top-down framework  in which gravity is an artifact of new small-scale physics,  it seems neither necessary nor appropriate to demand additional dynamics for the composite graviton. As an analogy, 
it would be incorrect to add dynamics for phonons to the microscopic description of the atomic physics responsible for the behavior of solids.

3) It is sometimes suggested that because regulators of ultraviolet divergences violate  axioms of relativistic quantum field theory, they cannot be physical. However, if quantum field theory is replaced with a different mathematical structure, the predictions of quantum field theory may be modified at small scales. For example, unitarity may be violated in a framework that does not include a Hilbert space of states among its axioms.

\subsection{Composite gravity}
From a bottom-up perspective, we are interested in a background-independent field theory without a fundamental dynamical spacetime metric. Such theories may be obtained by integrating out an auxiliary spacetime metric or vielbein minimally coupled to a field theory of interest. 
For a collection of $N$ scalar fields $\phi^a$ with potential $V(\phi^a)$, the composite metric operator is identified as \cite{Akama:1978pg,Akama:2013tua,Carone:2016tup},

 \begin{equation}
g_{\mu\nu}  \equiv \frac{\sum_a \partial_\mu\phi^a\,\partial_\nu\phi^a}{V(\phi^a)}
. \label{eq:compmetric} \end{equation} 
The resulting action for this toy model is (we take the speed of light $c=1$ from here on), \begin{equation}
S=\int d^4x\,\frac{\sqrt{\left|\det\sum_{a=1}^N\partial_\mu\phi^a\,\partial_\nu\phi^a\right|}}{V(\phi^a)}. \label{eq:action} \end{equation}

By expanding about a presumed expectation value $\langle g_{\mu\nu}\rangle$,
one can analyze the quantum theory as a theory in curved spacetime and determine self-consistently the condition for $\langle g_{\mu\nu}\rangle$ to equal the  metric of the curved spacetime. With $V(\phi^a)$ quadratic in the fields, and dimensional regularization as a regulator, it was shown that self-consistency is tantamount to vanishing of the expectation value of the energy-momentum tensor, which leads to Einstein's equations (plus corrections), with the gravitational coupling and cosmological constant determined by the regulator \cite{Batz:2020swk}. 

Then, with the same regulator in the self-consistent background, the four-point function $\langle \phi^a(x_1)\phi^b(x_2)\phi^c(x_3) \phi^d(x_4)\rangle$ was shown to contain a gravitational interaction \cite{Batz:2020swk}. The long-distance effective description includes Einstein gravity with cosmological constant and interactions determined by the potential $V(\phi^a)$.
This was presented as a toy model of composite gravity coupled to interacting matter, although to describe the theory completely we still require a physical regulator to replace dimensional regularization. 

\subsection{Stochastic composite gravity}
The suggestion that quantum theory might have an interpretation in terms of physical stochastic processes has a long history \cite{Fenyes,Doob:1942kiw,Bohm:1954zz,Nelson:1966,Nelson:1985,Guerra:1973ck,Guerra:1981ie,Parisi:1980ys,Barandes:2023ivl}. The application of stochastic processes to quantum gravity has witnessed a resurgence of interest from a variety of perspectives \cite{Erlich:2018qfc,Erlich:2024ooy,Moffat:1996fu,Huffel:1984mw,Kuipers:2021jlh,Kuipers:2021aok,Kuipers:2022wpy,Arzano:2024apl,Markopoulou:2003ps,Oppenheim:2018igd}. The main novelty of the present approach is that correlation functions are regularized by nonstandard kinematics.

Stochastic fields are distributions over parameters $x^\mu$, and across a stochastic event a field increases by a random local fluctuation drawn from a zero-centered Gaussian distribution. In a foliation with time coordinate $t$, 
the system is described kinematically by a discretized Itǒ process,
\begin{equation}d\phi^a(\mathbf{x})=b^a[\phi^a(\mathbf{x}),t]\,dt+\sqrt{2D_0}\,\int_t^{t+dt} dt'\,dW^a(x), \label{eq:dphi}
\end{equation}
where $b^a$ is a drift function, $D_0$ is a microscopic diffusion parameter, $dt$ is a sufficiently small time interval such that the expected change in $b^a$ is ${\cal O}(dt)$, and $dW^a$ describes the discrete random noise experienced by field $\phi^a$ such that, \begin{eqnarray}
\langle dW^a(x)\rangle&=&0, \nonumber \\ 
\langle dW^a(x)\,dW^b(x')\rangle&=&\delta^{ab}\,\sum_n\delta^4(x-x')\,\delta^4(x-x_n). \end{eqnarray}
Due to coordinate reparametrization invariance, the values of the coordinates $x_n$ of the stochastic events is arbitrary. 
The collection of stochastic events forms a causal set \cite{Bombelli:1987aa}. In particular, each event has a distinct past and future, which allows us to consider evolution of fields  toward the future in a coordinate-invariant context. From a semiclassical bottom-up perspective, the stochastic events are arranged in a Poisson sprinkling characterized by 
$\text{Probability(stochastic\ event/small\ unit 4-volume)}=1/\tau^4$, as in Fig.~1. 
\begin{figure}[t]
\begin{center}
\includegraphics[width = 0.2\textwidth]{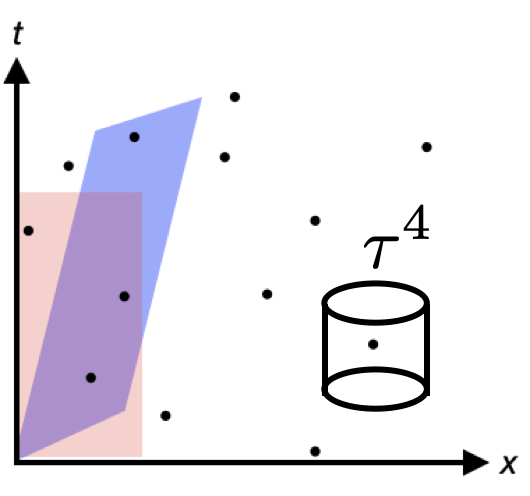}
\caption{Stochastic events are distributed in a Poisson sprinkling with characterstic 4-volume $\tau^4$. The Poisson distribution is Lorentz invariant, as indicated by the shaded volumes.}
\end{center}
\label{fig:Poisson}
\end{figure}
The 4-volume $\tau^4$ acts as a cutoff for the effective quantum field theory describing dynamics in the emergent spacetime geometry, and we note that  $\tau$ is invariant under Lorentz transformations.

Nelson showed that if a continuous stochastic process satisfies an appropriate equation of motion, then the probability distribution function associated with the process in stochastic equilibrium is determined by the Schrödinger equation and the Born rule \cite{Nelson:1985}. This description was generalized to QFT by Guerra and Ruggiero \cite{Guerra:1973ck,Guerra:1981ie}.

Conversely, given a solution to the Schrödinger equation, one can construct the corresponding stochastic process that in equilibrium has the quantum-mechanical probability distribution function and current. Writing the Schrödinger wavefunctional as
\begin{equation}
\Psi[\phi^a(\mathbf{x}),t]=e^{R[\phi^a(\mathbf{x}),t]+iS[\phi^a(\mathbf{x}),t]}, 
\label{eq:psiRS}\end{equation}
with real $R$ and $S$,  the corresponding stochastic process in a curved spacetime evolves according to  Eq.~(\ref{eq:dphi}) with drift coefficient,
\begin{equation}
b^a=\frac{\hbar}{\sqrt{|g|}\, |g^{00}|}\,\frac{\delta}{\delta\phi^a(\mathbf{x})}\left(S+R\right).
\label{eq:bRS}
\end{equation}

The functional Fokker-Planck equation, which describes the evolution of the probability density functional for the random fields, has an associated H-theorem which guarantees that generic ensembles of field configurations evolve to a common, possibly time-dependent, probability density function \cite{H-theorem}. We refer to the asymptotic solution of the Fokker-Planck equation as the equilibrium distribution.   It is analogous to the quantum equilibrium distribution of Bohmian mechanics.

 By an analysis similar to Nelson's for stochastic mechanics \cite{Nelson:1985}, we  conclude that generically, initial ensembles will evolve in $t$ towards a distribution satisfying the Born rule \begin{equation}
\rho[\phi^a(\mathbf{x}),t]=|\Psi[\phi^a(\mathbf{x}),t]|^2. \end{equation}
Planck's constant in the emergent quantum theory is related to the diffusion parameter $D_0$ and $\tau$ via $
\hbar/2=\tau^4\,D_0$.
Averaging over a 4-volume containing many stochastic events, we can approximate the discrete stochastic process by a continuous process with the same drift coefficients $b^a$, and with diffusion parameter \begin{equation}
D(x)=\frac{\tau^4\, D_0}{ \sqrt{|g|}\, |g^{00}|}. \end{equation}

In stochastic composite gravity, time derivatives in the composite metric operator are replaced by symmetrized combinations of forward and backwards stochastic derivatives \cite{Nelson:1985} defined by the expected value of differences in time towards the future $(\mathcal{D}_+\phi)$ or from the past $(\mathcal{D}_-\phi)$. \begin{eqnarray}
&g_{00}&=\frac{1}{2V(\phi^a)}\sum_a\left(\mathcal{D}_+\phi^a\,\mathcal{D}_+\phi^a +\mathcal{D}_-\phi^a\,\mathcal{D}_-\phi^a\right)%+{\cal O}(m^2\phi^2/V_0),
\nonumber  \\
&g_{i0}&=g_{0i}=\frac{1}{2V(\phi^a)}\sum_a\left( \mathcal{D}_+\phi^a+\mathcal{D}_-\phi^a\right)\partial_i\phi^a \nonumber \\
&g_{ij}&=\frac{1}{V(\phi^a)}\sum_a\partial_i\phi^a\,\partial_j\phi^a. \label{eq:Gmn}
\end{eqnarray}
As in the classical theory defined by the action Eq.~(\ref{eq:action}), the energy-momentum tensor vanishes in the stochastically quantized theory \cite{Erlich:2024ooy}. 
The canonical momenta have the form, \begin{equation}
\Pi^a_{\pm}=\frac12\sqrt{|g|}\frac{\left(g^{00}\,\mathcal{D}_\pm\phi^a+g^{0i}\partial_i\phi^a\right)}{V(\phi^a)},
\end{equation}
and satisfy Hamiltonian and momentum constraints:
\begin{equation}
\frac12\sum_a \left(\Pi_+^a\Pi_+^a+\Pi_-^a\Pi_-^a\right)-{\rm sign}(g)V(\phi^a)\det[g_{ij}]=0 \label{eq:Hconstraint}, \end{equation}
\begin{equation}
(\partial_i\phi^a)(\Pi_+^a+\Pi_-^a)=0, \label{eq:Pconstraint}\end{equation}
where $\det[g_{ij}]$ is the determinant of the 3-dimensional submetric $g_{ij}[\phi^a(\mathbf{x})]$.

In a semiclassical expansion we expand about the vacuum of the continuum curved-space quantum theory with a background metric $g^B_{\ \mu\nu}$. Expectation values of the Hamiltonian and momentum constraints will be satisfied in the semiclassical expansion as long as we can maintain vanishing of the expectation value of the energy-momentum tensor, which is analogous to the Virasoro constraints of perturbative string theory. 
Due to the stochastic discreteness, the nontrivial content of the constraints arise by replacing local fields by averages over small 3-volumes.  Equilibrium expectation values of these smoothed composite operators are finite due to the stochastic discreteness.

The result of all this can be understood from the bottom-up perspective, in which the stochastic discreteness serves as a covariant regulator for the composite gravity model. 
The analysis then mirrors Sakharov's induced gravity, demonstrating the emergent gravitational interaction with an emergent Planck mass \cite{Erlich:2024ooy} \begin{equation}
M_{{\rm Pl}}^2 \sim\hbar^2 N/\tau^2. \label{eq:MPl-stochastic}\end{equation}

\section{Predictions of Stochastic Composite Gravity}
\subsection{Bekenstein-Hawking Entropy Formula}
Gibbons and Hawking derived the Bekenstein-Hawking entropy formula, Eq.~(\ref{eq:Bek-Hawk}), by identifying the partition function of a gravitational theory with 
\begin{equation}
Z=e^{-(S_{EH}+S_{GHY})}, \label{eq:Z}\end{equation}
where $S_{EH}$ is the Euclidean Einstein-Hilbert action, and $S_{GHY}$ is the Gibbons-Hawking-York boundary term required by boundary conditions.
The Einstein-Hilbert action vanishes in the black-hole solution, but in the black-hole background the boundary term survives and reproduces the entropy law Eq.~(\ref{eq:Bek-Hawk}) with the correct factor of 1/4. For an excellent review of black-hole thermodynamics that benefits from the hindsight of recent developments, see Ref.~\cite{Witten:2024upt}. The interpretation of the Gibbons-Hawking result is initially puzzling: there is no {\em a priori} explanation for why the partition function $Z$ should have the form Eq.~(\ref{eq:Z}). We would expect the partition function to be determined by a trace over states, which in a field theory would correspond to a loop calculation, not a tree-level calculation.

Jacobson noted \cite{Jacobson:1994iw} that the Gibbons-Hawking calculation, or an equivalent calculation by Susskind and Uglum \cite{Susskind:1994sm}, would describe both the entaglement entropy and the thermal entropy of a black hole in the setting of Sakharov's induced gravity. The reason is simply that the gravitational action arises from the regularized divergences that appear when integrating out the matter fields, which maps precisely to the leading contribution to the partition function that determines the entropy.  
What was lacking in this story was a physical regulator responsible for the induced gravitational interaction. Here, the regulator is provided by the discreteness of the stochastic processes describing states of the system, with the corresponding Planck mass given by Eq.~(\ref{eq:MPl-stochastic}).

This also explains the species puzzle for the black-hole entropy formula, which is the question of why the black hole entropy does not appear to depend on the number of matter species. The explanation in our scenario is that the induced gravitational constant $G_N\propto1/M_{{\rm Pl}}^2$ scales inversely with the number of species according to Eq.~(\ref{eq:MPl-stochastic}), precisely accounting for the expected scaling of the entropy Eq.~(\ref{eq:Bek-Hawk}) with number of species.

\subsection{Early Cosmology and the CMB}
The pilot wave description of quantum field theory is a cousin of the stochastic framework. Both descriptions include the values of the fields in their ontologies, but the pilot wave  lacks an obvious physical ultraviolet regulator. Valentini noticed that in a pilot wave description of the inflaton field during inflation,  the system would naturally be out of equilibrium initially, which would lead to differences from the standard predictions of the power spectrum of the cosmic microwave background (CMB) \cite{Valentini:2008dq,Valentini:2015sna}. In particular, out-of-equilibrium effects would be most apparent for longer wavelength modes, naturally leading to a reduction in power in the low-$\ell$ moments of the CMB power spectrum. Interestingly, this is consistent with data from the Planck experiment \cite{Planck:2018nkj}, although the low-$\ell$ moments are not well predicted due to cosmic variance. Here we just note that a similar effect occurs in the stochastic composite gravity framework, as will be demonstrated elsewhere \cite{Delgado-Erlich-Staker}. It is interesting to also speculate about implications for late-time cosmology and the interpretation of dark energy.

\subsection{Euclidean Description at Ultrashort Distances}

As noted by Guerra and Ruggiero \cite{Guerra:1973ck,Guerra:1981ie}, there is a relation between the stochastic process describing the relativistic vacuum and the corresponding Euclidean theory.
For example, in flat spacetime, the ground state of a free scalar field is described formally by the process \cite{Guerra:1973ck},
\begin{equation}
d\phi(\mathbf{x})=-\sqrt{-\nabla^2+m^2}\,\phi(\mathbf{x})\,dt+\sqrt{2D_0}\,\int dt\,dW(x).
\label{eq:dphi_Mink}\end{equation}
Away from stochastic events, $dW(x)=0$. Hence, over small scales, Eq.~(\ref{eq:dphi_Mink}) can be rewritten,
\begin{equation}
\frac{\partial\phi}{\partial t}=-\sqrt{-\nabla^2+m^2}\,\phi. \label{eq:dW0}\end{equation}
A $t$-derivative of Eq.~(\ref{eq:dW0}) yields the Euclidean field equation. The relation between the Lorentzian QFT and the Euclidean stochastic process makes the semiclassical expansion and the Hamiltonian constraint  a bit subtle.
It is tempting to speculate that in a cosmological context, the singularity at early times is replaced by a description that is both Euclidean and classical, by analogy with the ultrasmall-scale behavior of the free theory in the stochastic description. This is reminiscent of the Hartle-Hawking no-boundary proposal \cite{Hartle:1983ai}, and would clarify the nature of the transition from the Euclidean to Lorentzian signature as occuring in conjunction with a transition from classical to stochastic behavior.

\subsection{The Arrow of Time}
As Peierls \cite{Peierls}, Penrose \cite{Penrose:Emperor,Penrose:Road}, and others have emphasized, the early universe, homogeneous over scales that grew to become our observable universe, appears to have been in an exceptionally low entropy state. In the stochastic composite gravity setting, generic initial conditions would be far from stochastic equilibrium. Gravitation would not yet be active because the quantum field theory from which gravity emerges would not yet be an accurate description. 
The system would evolve towards a high-entropy state which, in the absence of gravity, may be homogeneous except for stochastic fluctuations, as required for the initial conditions for inflation. 
In this scenario, the late-time behavior of a closed universe would involve large field fluctuations over small scales, and once again gravity would cease to be a valid description. The symmetry between conditions of the universe at early and late times may be restored due to the absence of gravity in the conditions of the early and late universe.

\subsection{Experimental tests}
Out of stochastic equilibrium, the predictions of the stochastic field theory and quantum field theory disagree. One can envision sudden transitions of quantum states, for example during measurement of atomic transitions, during which the system may be considered temporarily out of equilibrium. By analyzing time scales associated with the approach to equilibrium, which depends on the system, it may be possible to design atomic physics experiments involving fast laser pulses or precision measurements to test the stochastic framework against predictions of standard quantum theory. In a similar spirit, it is possible that collider data, which provides plentiful agreement with the predictions of quantum field theory, might also provide tests of the stochastic framework.

\section{Conclusions}
This essay described a fundamental framework in which both quantum field theory and general relativity are emergent descriptions of a system of stochastic fields. The framework resolves a number of puzzles, makes a number of predictions, and may be experimentally testable. There is no obvious obstruction to extending the framework to include fermions and gauge fields, in analogy with the stochastic description of QCD in the framework of Parisi and Wu \cite{Parisi:1980ys,Damgaard:1983tq,Damgaard:1987rr}. However, this has not yet been done in the setting of stochastic composite gravity.

%\end{document}

\pagebreak

\end{document}